\documentclass[prb,twocolumn,superscriptaddress,amsmath,amssymb,amsthm]{revtex4}

\usepackage{graphicx,epstopdf,hyperref}
\usepackage{amsmath,amssymb,amsthm,amsfonts}
\usepackage{color}
\usepackage{bm}

\definecolor{MS-color}{RGB}{128,0,128}

\begin{document}



\title{Self-consistent calculation of the flux-flow conductivity in diffusive superconductors}

\author{A. Vargunin}
\affiliation{Department of Theoretical Physics, The Royal Institute of
Technology, Stockholm, SE-10691 Sweden}
\affiliation{Institute of Physics, University of Tartu, Tartu, EE-50411,
Estonia}
 \author{M. A.~Silaev}
 \affiliation{Department of
Physics and Nanoscience Center, University of Jyv\"askyl\"a, P.O.
Box 35 (YFL), FI-40014 University of Jyv\"askyl\"a, Finland}
\date{\today}

\begin{abstract} 
 In the framework of Keldysh-Usadel kinetic theory, we study the temperature dependence of 
flux-flow conductivity (FFC) in diffusive superconductors.
By using self-consistent vortex solutions we find the exact values of dimensionless parameters that determine
 the diffusion-controlled FFC both in the limit of the low temperatures and close to the critical one. 
Taking into account the electron-phonon scattering we study the transition between flux-flow regimes 
controlled either by the diffusion or the inelastic relaxation of non-equilibrium quasiparticles.
 We demonstrate that the inelastic electron-phonon relaxation leads to the strong suppression of FFC 
 as compared to the previous estimates making 
it possible to obtain the numerical agreement with experimental results. 
 \end{abstract}

 \maketitle

 \section{Introduction}
  Vortex motion is an important process that determines resistive properties of type-II superconductors in the flux-flow regime. 
  At magnetic fields $B$ much weaker then the upper critical one, $H_{c2}$, 
  the density of vortex lines is small and the total electric losses are given by the superposition
  of the individual vortex contributions. In this regime, the flux-flow resistivity $\rho_f$ is proportional to the density of vortex lines, 
  $\rho_f\sim B/H_{c2}$,
  as described by the general expression suggested by Bardeen and Stephen \cite{BS}. 
  The inverse quantity $\sigma_f= 1/\rho_f$, the flux-flow conductivity (FFC)  is therefore given by  
  \begin{equation} \label{Eq:BetaDefinition}
  \sigma_f/\sigma_n= \beta H_{c2} / B,
  \end{equation}
  where $ \sigma_n$ is the normal state conductivity and $\beta$ is the numerical coefficient which is determined by 
  the particular microscopic model. 

For superconducting materials with high rate of impurity scattering, 
 the numerical value of $\beta\approx0.9$ at low temperatures has been reported by Gor'kov and Kopnin \cite{GorkovKopnin1}(GK).
  This value of the dimensionless parameter has been obtained using the approximate vortex solution. 
Up to date the exact value of $\beta$ has been unknown and it is reported in the present paper based on the fully-self consistent 
vortex structure calculations.  

At elevated temperatures, two different regimes of the vortex motion has been considered, depending on the 
dominant mechanism of the relaxation\cite{kopnin-book}. One of them is the diffusion-controlled flux-flow,
when the generation of non-equilibrium quasiparticles near the vortex line is balanced by their diffusion to the infinity. 
As temperature approaches $T_c$
this mechanism 
results in the divergent behaviour of FFC given by\cite{GorkovKopnin2,LarkinOvchinnikovTc2,kopnin-book}:
 \begin{equation} \label{EqBetaDiverge}
 \beta\approx \beta_0(1-T/T_c)^{-1/2}
 \end{equation}
 with the temperature-independent $\beta_0$.  
Qualitatively this behaviour is explained by the  vortex core size increase proportional to the Ginzburg-Landau coherence length 
$\xi_{GL}(T) \sim \sqrt{D/ (T_c-T)}$, where $D$ is the diffusion coefficient. 
 This dependence is in the qualitative agreement with experimental results \cite{Gorkov1975} pointing to 
 the significant increase of $\beta$ as temperature approaches $T_c$. 
 However, quantitative agreement is lacking. Initially, the value of $\beta_0 \approx 1.1$ has been reported\cite{GorkovKopnin2} 
 which by coincidence was in the good agreement with experiments\cite{Gorkov1975}. 
 However subsequently this result has been revised to $\beta_0 \approx 4.04$ {by Larkin and Ovchinnikov \cite{LarkinOvchinnikovTc2,kopnin-book} (LO)} which is  several times larger than the  measured values in 
 various superconductors \cite{ExperimentKim,ExperimentVinen,ExperimentGoncharov,ExperimentMuto,ExperimentTakayama,ExperimentFogel,poon}.
 
When the temperature becomes sufficiently close to $T_c$ the relaxation 
is dominated by the inelastic electron-phonon collisions. 
 This regime is described by generalized time-dependent Ginzburg-Landau theory {(GTDGL)}
  yielding the FFC decreasing with temperature \cite{kopnin-book}
  \begin{equation} \label{Eq:BetaTDGL}
  \beta \sim (T_c\tau_{ph}) (1-T/T_c)^{1/2},
  \end{equation}
  where $\tau_{ph}$ is the electron-phonon relaxation time. 
 In the limit $T\to T_c$ the gappless superconducting state is realized. 
 In this case 
 the decrease of $\beta(T)$ saturates at $\beta=1.45$ \cite{Schmid}. 
 
  The crossover between two regimes described by the Eqs.(\ref{EqBetaDiverge},\ref{Eq:BetaTDGL})
  occurs at the temperatures $\tau_{ph}(T_c-T) \sim 1$  when the diffusion rate becomes of the order of electron-phonon relaxation rate, 
  $D\xi_{GL}^{-2} \sim \tau_{ph}^{-1}$. That yields an estimation of the maximal value 
   $\max(\beta) \sim \sqrt{T_c\tau_{ph}} $ obtained from Eqs. (\ref{EqBetaDiverge}) and (\ref{Eq:BetaTDGL}) at the upper and lower 
  borders of their applicability respectively. 
   
 Although the estimations of $\max (\beta)$ obtained from Eqs. (\ref{EqBetaDiverge}) and (\ref{Eq:BetaTDGL}) agree by the order of magnitude,
  the temperature domains where these equations are valid do not overlap. 
  Therefore, to find the behaviour of $\beta$ 
 in the transition interval 
 from the diffusion-controlled to the
  GTDGL regime one needs to improve the accuracy of the calculation taking into account both mechanisms of relaxation.
  This is the problem we address in the present paper. We study the linear response FFC of the 
  sparse vortex lattices in small magnetic field  by solving numerically 
   kinetic equations describing non-equilibrium states generated by moving isolated vortices. 
   To find kinetic coefficients and driving terms we use vortex structures calculated self-consistently.

For the diffusion-controlled vortex motion, we calculate the temperature dependence $\beta= \beta(T)$ 
 and compare it with the interpolation curve suggested in earlier works\cite{kopnin-book,Gorkov1975}.
Taking into account the electron-phonon scattering, we demonstrate that it
leads to the significant suppression of the FFC at intermediate temperatures, $\tau_{ph}(T_c-T)\sim 1$, as compared to the estimations obtained from the Eqs.(\ref{EqBetaDiverge},\ref{Eq:BetaTDGL}). Using the electron-phonon  relaxation rate $\tau_{ph}^{-1}$ as the fitting parameter 
we obtain numerically accurate fits to the experimentally measured temperature dependencies of FFC 
in Zr$_3$Rh \cite{poon} and Nb-Ta \cite{ExperimentKim,ExperimentVinen}.

The structure of this paper is as follows. In Sec.\ref{Sec:KinEq} we introduce the Keldysh-Usadel description of the 
kinetic processes in dirty superconductors. Here the basic components of the kinetic theory are discussed including kinetic equations, 
self-consistency equations for the order parameter and the general expression for the viscous force acting 
 on moving vortices. Sec. \ref{Sec:theta} introduces $\theta$-parametrization of the theory.
  Calculated {temperature} dependencies of FFC are reported in Sec.\ref{Sec:results} for different regimes. 
 {The diffusion-controlled flux-flow is discussed in Sec.(\ref{Sec:DiffusionControlled}) and the influence of increasing 
 electron-phonon relaxation rate} is studied in   Sec.(\ref{Sec:InelasticRelaxation}). 
     The work summary is given in Sec. \ref{Sec:Conclusion}.

 \section{Kinetic equations and the forces acting on the moving vortex line}\label{Sec:KinEq}
 The quasiclassical Green's function (GF) is defined as
 \begin{equation}
 \check{g} = \left(%
 \begin{array}{cc}
  \hat g^R &  \hat g^K \\
  0 &  \hat g^A \\
 \end{array}\label{eq:GF0}
 \right)\; ,
 \end{equation}
 where $g^K$ is the (2$\times$2 matrix) Keldysh component 
 and $\hat g^{R(A)}$ are the retarded (advanced) ones. 
  The GF $\check{g} = \check{g}  (t_1,t_2,{\bm r})$ depends on times $t_{1,2}$ and a single spatial coordinate ${\bm r}$. 
 In dirty superconductors  $\check g$ obeys the Keldysh-Usadel equation 
 \begin{equation}\label{Eq:KeldyshUsadelText}
 \{\hat\tau_3\partial_t, \check g \}_t = D\hat\partial_{\bf r} 
 ( \check g \circ \hat\partial_{\bf r} \check g) + [\hat H, \check g]_t  + \check I .
 \end{equation} 
 Here 
   $\hat\tau_{0,1,2,3}$ are Pauli matrices in Nambu space, $D$ is the diffusion constant,   
 $\hat H ({\bm r},t) = i \hat \Delta{-i e \phi\hat\tau_0}$, where $\hat\Delta (t)= i |\Delta| \hat\tau_2 e^{-i\varphi \hat\tau_3}$ 
 is the gap operator, $\varphi$ is the gap phase and $\phi$ is the electrostatic potential.
 In Eq.(\ref{Eq:KeldyshUsadelText}) the commutator is defined as  
   $[X, g]_t= X(t_1) g(t_1,t_2)- g(t_1,t_2) X(t_2)$, similarly for anticommutator 
   $ \{X, g\}_t= X(t_1) g(t_1,t_2)+ g(t_1,t_2) X(t_2)$. The symbolic product operator is given by
   $ (A\circ B) (t_1,t_2) = \int_{-\infty}^{\infty} dt A(t_1,t)B(t,t_2)$ and the covariant differential superoperator is
   \begin{align}
  \hat \partial_{\bf r} \check g= \nabla \check g -ie \left[\hat\tau_3{\bm A}, \check g \right]_t .
  \end{align}
  
  The collision integral in (\ref{Eq:KeldyshUsadelText}) is given by 
  \begin{equation} \label{Eq:SelfEnergy}
  \check I =  i (\check g\circ \check \Sigma - \check \Sigma\circ\check g),
  \end{equation}
  where the self energy $\check \Sigma$ may contain contributions from different relaxation processes. 
  Here we take into account only the electron-phonon scattering which plays an important 
  role in the energy relaxation. 
     
   The Keldysh-Usadel Eq. (\ref{Eq:KeldyshUsadelText}) is complemented by 
   the normalization condition
   $(\check g\circ \check g) (t_1,t_2)= \check \delta (t_1 - t_2)$
   which allows to introduce parametrization of the Keldysh component in 
   terms of the distribution function
   \begin{eqnarray}
   \hat g^K (t_1,t_2) = (\hat g^R \circ \hat f)(t_1,t_2) - (\hat f\circ \hat g^A)(t_1,t_2),  
   \label{Eq:Parametrization}
   \\
   \hat f(t_1,t_2)= \hat\tau_0 f_L (t_1,t_2) + \hat\tau_3 f_T (t_1,t_2). 
   \label{Eq:DistrFun}
   \end{eqnarray} 
   The deviation of $f_L$ from the equilibrium distribution is related to the effective temperature change, 
   and $f_T$ is the charge imbalance on the quasiparticle branch.
 
 To proceed we introduce the mixed representation in the time-energy domain as follows
  $ \check g(t_1,t_2) = \int_{-\infty}^{\infty} \check g(\varepsilon, t) 
  e^{-i\varepsilon(t_1-t_2) } d\varepsilon/(2\pi) $, 
  where $t=(t_1+t_2)/2$.  
  The Keldysh-Usadel equation (\ref{Eq:KeldyshUsadelText}) can be simplified by using the 
  gradient approximation. In order to keep the 
  resulting kinetic equations gauge invariant we use the modified GFs 
  $ \check g_{new}(t_1,t_2) = \hat W (t_1 ,t) \check g (t_1,t_2) \hat W (t ,t_2) $  
  where the link operator is given by $\hat W (t_1 ,t_2) = e^{i \hat\tau_3 \int_{t_1}^{t_2} e \phi dt }  $.
  This transformation leads to the local chemical potential shift.
  To take this into account  we will use the substitution $f_T (\varepsilon ,t)\to f_T (\varepsilon ,t){+e\phi \partial_\varepsilon f_0} $,
  where $f_0 (\varepsilon) = \tanh (\varepsilon/2T) $ is the equilibrium distribution function.
  After this transformation $f_T (\varepsilon ,t)$ denotes the deviation from the local equilibrium distribution.      
   
    Then, keeping the first order non-equilibrium terms 
  we obtain the system of two coupled kinetic equation that determine both the transverse 
  and longitudinal distribution function components $f_{L,T} = f_{L,T} (\varepsilon,t)$ 
  (the detailed derivation is given in the Appendix\ref{App:KinDerivation})
   \begin{align} \label{Eq:KineticEqFT3} 
   & \nabla ( {\cal D}_T \nabla f_{T} )+ 
 {\bm j}_e\cdot\nabla f_L +  2i {\rm Tr} [ (\hat g^R + \hat g^A) \hat \Delta] f_T = \nonumber\\ 
 & \partial_\varepsilon f_0 {\rm Tr} [ \hat\tau_3 \hat \partial_t\hat \Delta ( \hat g^R+ \hat g^A) ]  ,\\
\label{Eq:KineticEqFL} 
 &\nabla ( {\cal D}_L \nabla f_{L}) + {\bm j}_e\cdot\nabla f_T + 2i {\rm Tr}
 [\hat\tau_3 (\hat g^R -\hat g^A) \hat \Delta] f_T =  \nonumber\\  
 & - \partial_\varepsilon f_0 {\rm Tr}[ \hat \partial_t\hat \Delta( \hat g^R - \hat g^A) ] 
 -{{\rm Tr}\hat J} , 
 \end{align} 
where the energy-dependent diffusion coefficients ${\cal D}_{T,L}$ and the spectral charge current ${\bm j}_e$ are given by 
 \begin{align} \label{Eq:DiffCoeffT} 
& {\cal D}_T = D{\rm Tr}(\hat\tau_0 - \hat\tau_3\hat g^R \hat\tau_3\hat g^A), \\
 \label{Eq:DiffCoeffL}
 &{\cal D}_L =  D{\rm Tr} (\hat\tau_0 - \hat g^R \hat g^A ), \\
 \label{Eq:SpectralCurrent}
 &{\bm j}_e =  D{\rm Tr}\; [ \hat\tau_3( \hat g^R\hat \nabla \hat g^R - \hat g^A \hat\nabla \hat g^A)].   
 \end{align}
 In Eqs. (\ref{Eq:KineticEqFT3},\ref{Eq:KineticEqFL},\ref{Eq:SpectralCurrent}) we use the covariant time derivative and 
 spatial gradient defined by $\hat \partial_t = {\hat\tau_0\partial_t+ 2i e\phi \hat\tau_3}$ and 
 $ \hat\nabla = \nabla - ie{\bm A} [\hat\tau_3, ] $.
We omit the driving terms containing electric field which is justified in type-II superconductors with large Ginzburg-Landau parameters.
In such systems the dominating  driving terms are those containing order parameter gradients.  
 
 The electron-phonon collision integral in the r.h.s. of kinetic equation (\ref{Eq:KineticEqFL}) 
 is $\hat J=\hat I^K-\hat I^R\circ\hat f+\hat f\circ\hat I^A$, where 
 the components of $\check I$ are given by Eq.(\ref{Eq:SelfEnergy})  with electron-phonon self-energies
 \cite{eliashberg72} 

\begin{align}\label{Eq:ElectronPhononSE}
 &  \hat  \Sigma^{R/A/K} (\varepsilon) =  -\frac{\lambda_{ph}}{56\zeta(3)T_c^2}  \int_{-\infty}^{\infty} 
 d\omega \tilde{\Sigma}^{R/A/K}(\omega,\varepsilon), 
 \\
 & \tilde{\Sigma}^{R/A}(\omega,\varepsilon) =  D^K(\omega)\hat g^{R/A} (\varepsilon+\omega) -D^{R/A}(\omega)\hat g^K (\varepsilon+\omega), \nonumber\\
 & \tilde{\Sigma}^K(\omega,\varepsilon) = D^K(\omega)\hat g^K (\varepsilon+\omega) - D^{RA}(\omega) \hat g^{RA} (\varepsilon+\omega) .
 \nonumber
\end{align}
 Here 
 \begin{align}
 & D^{R/A}(\omega)=\pm i\omega |\omega|,
  \\
 & D^K(\omega) = D^{RA}(\omega)\coth\left(\frac{\omega}{2T}\right)
 \end{align}
 are the free phonon propagators, $D^{RA}= D^R - D^A $ and 
 $\hat g^{RA}= \hat g^R - \hat g^A $.
  We parametrise the electron-phonon self-energy by dimensionless constant 
  $\lambda_{ph} = (T_c\tau_{ph})^{-1}$ where $\tau_{ph}$ is electron-phonon relaxation time at $T=T_c$.

 The force acting on the moving vortex line 
 from the dissipative environment can be calculated 
 according to the expression \cite{LarkinOvchinnikov,kopnin-book} 
  \begin{equation} \label{Eq:FenvGen}
 {\bm F}_{env} = \nu \int d^2 {\bm r} \int_{-\infty}^{\infty}
 \frac{d\varepsilon}{4} 
 {\rm Tr} (\hat g^{nst} \hat \partial_{\bm r} \hat \Delta). 
  \end{equation}
 where $\nu$ is the density of states and  $\hat g^{nst}$ is the non-stationary Green's function which can be obtained by
  the gradient expansion as follows 
 \begin{align}\label{Eq:GnstExp}
 &\hat g^{nst}= -\frac{i}{2} {\hat\partial_t} (\hat g^R + \hat g^A)\partial_\varepsilon f_0 + 
 \\
 & (\hat g^R - \hat g^A )(f_L-f_0) +  (\hat g^R\hat\tau_3 - \hat\tau_3\hat g^A)f_T.
 \nonumber
 \end{align} 
Here $f_T$ denotes the deviation from the local equilibrium as discussed above.
 In Eq.(\ref{Eq:FenvGen}) we neglect the contribution from the normal component of the charge current. 
 This assumption is well justified for the small magnetic fields as compared to the upper critical one\cite{Silaev2016}.       
       
 \section{$\theta$-parametrization}\label{Sec:theta}
 
 In general, the normalization condition allows one to parametrize GF by complex variables
  $\theta$ and $\tilde\varphi$. For the axially symmetric vortices the latter coincides with the vortex phase $\tilde{\varphi} = \varphi$.
  In this case  we have
 \begin{align}\label{Eq:SpectralGF-R}
 & \hat g^R = \hat\tau_3\cosh\theta + i\hat\tau_2 e^{-i\hat\tau_3 \varphi} \sinh\theta,
 \\ \label{Eq:SpectralGF-A}
 & \hat g^A =-\hat\tau_3\cosh\theta^\ast - i\hat\tau_2 e^{-i\hat\tau_3 \varphi} \sinh\theta^\ast.
 \end{align}
 The complex parameter $\theta = \theta(r)$, depending only on the distance to the vortex centre $r$ 
  is given by the solution of the Usadel equation
  \begin{align}\label{Eq:theta}
 \nabla^2_r\theta - \frac{\sinh 2\theta}{2r^2} + 
 \frac{2i}{D} \left[ \left(\varepsilon + \frac{i}{2\tau}\right) \sinh\theta - |\Delta|\cosh\theta\right] = 0 , 
 \qquad
 \end{align}
 where $\nabla_r^2 = \partial_r^2 + r^{-1} \partial_r $, see Appendix \ref{App:theta}. 
 The boundary conditions for Eq.(\ref{Eq:theta}) read 
 \begin{align}\label{thetabc}
 & \theta(0)=0, \\ 
 & \sinh\theta(\infty)= \Delta_0/ \sqrt{ [\varepsilon + i/(2\tau)]^2 - \Delta_0^2 },
 \end{align}
 where $\Delta_0 = |\Delta(r=\infty)|$.
Electron-phonon scattering with characteristic time $\tau$ in Eqs.(\ref{Eq:theta},\ref{thetabc}) 
 regularizes spectral functions near the gap edge singularity.
 At low temperatures electron-phonon scattering 
 does not affect the calculation results. 
 In the vicinity of $T_c$, its value is important since the inelastic relaxation dominates the dissipation. To describe the effects of electron-phonon scattering on the relaxation we
  calculate $\tau$ self-consistently  within the relaxation-time approximation described in Appendix \ref{App:coll}. In this approach
 \begin{align}\label{tau_theta}
 \frac{1}{\tau} =\frac{\lambda_{ph} \cosh\frac{\varepsilon}{2T}}{14\zeta(3)T_c^2} 
 \int_{-\infty}^{\infty} \frac{ \omega|\omega| d\omega {\rm Re}\cosh[\theta(\varepsilon+\omega)] }{\sinh\frac{\omega}{2T} \cosh\frac{\varepsilon + \omega}{2T}}.
 \end{align}

   To determine the gap profile, we use a stationary self-consistency equation written in the form  
  \begin{equation}\label{Eq:SelfConsistency2B-2}
  |\Delta|\ln(T/T_c)=2\pi T 
  \sum_n \left( \sin \theta^M_n - |\Delta|/\omega_n\right).
  \end{equation}
  Here the summation runs over Matsubara frequencies $\omega_n =(2n+1)\pi T$, $n=0\ldots\infty$, and the angle $\theta^M_n(r)$ 
  parametrizes imaginary-frequency GF obtained by the transformation $\theta\to-i\theta^M_n$ 
  from the Eq.(\ref{Eq:SpectralGF-R},\ref{Eq:SpectralGF-A}) in the upper and lower half-planes, respectively.    
  To obtain $\theta^M_n(r)$ we solve the Usadel equation (\ref{Eq:theta}) with $\theta\to-i\theta^M_n$ and 
  $\varepsilon\to i\omega_n$. 
We assume that the condition $\omega_n \tau \gg 1$ is always satisfied and neglect relaxation time correction while solving 
  Eq. \ref{Eq:theta} for the imaginary frequencies. 
  The boundary conditions read as 
  $\theta^M_n(0)=0$ and
  \begin{align}\label{Bc-Usadel}
  \theta^M_n(\infty)= \sin^{-1}\left[ \Delta_0/
  \sqrt{\Delta^2_0+\omega_n^2} \right]. 
  \end{align}

The driving terms 
 in kinetic equations (\ref{Eq:KineticEqFT3},\ref{Eq:KineticEqFL})
are given by the time-derivatives of the order parameter which for the steady vortex motion
 can be written as $\partial_t \hat{\Delta} = -{\bm v}_L\cdot\nabla \hat{\Delta} $,
 where ${\bm v}_L$ is the vortex velocity. For the axially-symmetric vortex, 
this form of driving terms allows for the separation of the variables using the ansatz 
 \begin{align} \label{Eq:DFansatz}
 & f_L -f_0 =  v_L\tilde{f}_L \partial_\varepsilon f_0 \cos\varphi \\
 & f_T = v_L\tilde{f}_T  \partial_\varepsilon f_0 \sin\varphi .
 \end{align}  
 Here the amplitudes $\tilde{f}_{L,T} = \tilde{f}_{L,T} (r) $ are given by the ordinary differential equations
which can be written in the compact form as follows
  \begin{align} \label{Eq:tildefT}
  &\partial_r(r\mathcal{D}_T\partial_r\tilde f_T)  -\left( \mathcal{D}_T - 8|\Delta|r^2\cosh\vartheta \sin \eta  \right) \frac{\tilde f_T}{r} = \\ 
  &\qquad 4|\Delta|\cosh\vartheta\sin \eta - \mathcal{D}_L\sinh(2\vartheta)\tan\eta \frac{\tilde f_L}{r},\nonumber\\
   & \partial_r(r\mathcal{D}_L\partial_r\tilde f_L) - \mathcal{D}_L\frac{\tilde f_L}{r}  =  \mathcal{D}_T\tanh\vartheta \sin(2\eta) \frac{\tilde f_T}{r} -
   \label{Eq:tildefL}
   \\
   &\qquad  4r\sinh \vartheta\cos\eta \partial_r|\Delta| 
 +r \tilde f_L\nu_{out}-rj_{in},\nonumber
   \end{align}   
where $\vartheta = {\rm Re}\theta$ and $\eta = {\rm Im}\theta$.
For detailed derivation see Appendix \ref{App:theta}. The last two terms in Eq.(\ref{Eq:tildefL})
describe scattering-out and scattering-in contributions to the inelastic relaxation of the non-equilibrium longitudinal imbalance. The integrals are given by  	
\begin{align}\label{collint_out1}
&\nu_{out}=\frac{2\lambda_{ph}\cos\eta}{7\zeta(3)T_c^2}\int_{-\infty}^\infty d\omega\omega|\omega|\cos[\eta(\varepsilon+\omega)]\times\nonumber\\
&\cosh[\vartheta(\varepsilon)-\vartheta(\varepsilon+\omega)]\left[1/f_0(\omega)-f_0(\varepsilon + \omega)\right],\\
&j_{in}=\frac{2\lambda_{ph}\cos\eta}{7\zeta(3)T_c^2\partial_{\varepsilon}f_0}\int_{-\infty}^\infty d\omega\omega|\omega|\cos[\eta(\varepsilon+\omega)]\tilde f_L(\varepsilon + \omega)\times\nonumber\\
&\cosh[\vartheta(\varepsilon)-\vartheta(\varepsilon+\omega)]\partial_{\varepsilon}f_0(\varepsilon + \omega)[f_0(\varepsilon)+1/f_0(\omega)]. \label{collint_in1}
\end{align}

Kinetic equations (\ref{Eq:tildefT},\ref{Eq:tildefL}) are solved numerically within the interval $0 \leq r \leq r_c$, where $r_c$ is the cell radius. For the regime of diffusion-controlled dissipation we choose the interval large enough 
so that the result is not sensitive to $r_c$. When discussing the crossover to the inelastic relaxation-driven 
dissipation we set the interval to be larger than the inelastic relaxation length, $\sqrt{D\tau}$, which determines the decay of 
$\tilde f_L$ at large distances. 
  We use the following boundary conditions
 \begin{align} \label{Eq:BCzero}
 & \tilde{f}_T (r =0) = \tilde{f}_L (r =0) =0,
 \nonumber 
 \\ 
 & \tilde{f}_T (r =r_c)  = 1/2r_c , 
 \\ 
 & \partial_r\tilde f_L (r =r_c) = 0.
 \nonumber 
 \end{align}
 Here the condition at $r=0$ in  Eq.(\ref{Eq:BCzero}) follows from the regularity of the solutions at the origin, 
 while condition  at $r_c$ provides the disappearance of charge imbalance and the absence of the heat flow into the bulk.
  
 The viscous friction force acting on individual 
 moving vortex can be written as ${\bm F}_{env} = -\varrho {\bm v}_L$.
  We present viscosity coefficient in the form $\varrho= \pi\hbar\nu (\alpha+ \gamma)$ separating
 the contributions of the driving terms related to the 
   gap modulus and phase gradients, see Appendix \ref{App:theta}.
In general the flux-flow conductivity can be expressed through the vortex
 viscosity  as follows \cite{GorkovKopnin2}
 \begin{equation}\label{Eq:FluxFlowCond}
 \sigma_f= \varrho /(B\phi_0),
 \end{equation}
 where $\phi_0$ is the magnetic flux quantum.
 Taking into account the normal-state Drude conductivity, $\sigma_n=2e^2\nu D$, we write the  
 FFC in the form (\ref{Eq:BetaDefinition}) with 
 \begin{equation} \label{Eq:FluxFlowCondVisc}
  \beta= c(\alpha + \gamma)/(2eDH_{c2}).
 \end{equation}
 The upper critical field $H_{c2}$ is determined by the Maki equation \cite{Maki69}, $\ln(T/T_c)+\psi(1/2+eH_{c2}D/(2\pi cT))=\psi(1/2)$, where $\psi$ is digamma function. The low-temperature limit gives $H_{c2}=\phi_0T_c/(2D \gamma_0)$, where $\gamma_0=1.781$. Close to the critical temperature one obtains $H_{c2}=\phi_0/(2\pi\xi_{GL}^2)$  and $\xi_{GL} = \sqrt{\pi D\hbar/8(T_c-T)}$ is the Ginzburg-Landau correlation length.

\section{Results} \label{Sec:results}  

\subsection{Diffusion-controlled flux flow} \label{Sec:DiffusionControlled}  

 When the temperature is sufficiently far from the critical one the electron-phonon relaxation terms
 in the kinetic equation (\ref{Eq:tildefL}) can be neglected being much smaller than the diffusion one. 
 Qualitatively this approximation means that the non-equilibrium quasiparticles generated near the vortex
 can drift to the infinity
 at the rate exceeding the one of inelastic relaxation. This regime is called the 
 diffusion-controlled flux-flow and it is realized in the temperature domain $(T_c-T)\tau_{ph} \gg 1$.
 {Below we analyse this scenario separately for different temperature intervals.}

 \subsubsection{Low-temperature limit}
  At low temperatures, the sizeable quasiparticle density exists only inside vortex cores where the superconducting 
  order parameter is suppressed.  
  In this case, it is sufficient to consider only zero-energy GF for which parameter $\theta$ is purely imaginary, 
  $\vartheta=0$. 
 The dissipation is dominated by the charge relaxation processes described by the distribution function $f_T$. 
  The effective temperature change described by the distortion of $f_L$ can be neglected.    
  Then Eq. (\ref{Eq:tildefT}) can be written as follows 
   \begin{align} 
   \label{Eq:tildefTlowT}
   & \nabla_r^2 \tilde f_T - \left(\frac{1}{r^2}-\frac{2|\Delta|}{D} \sin\eta\right)\tilde f_T = 
   \frac{|\Delta|}{Dr}\sin\eta.
   \end{align} 
 As a result, the coefficients that determine FFC in Eq.(\ref{Eq:FluxFlowCondVisc}) are given by 
  \begin{align} \label{Eq:VisclowT}
  & \alpha =- \int_0^\infty rdr \partial_r|\Delta|\partial_r\sin\eta, \\
  & \gamma = \int_0^\infty dr |\Delta|(2\tilde f_T-1/r)\sin\eta.\nonumber   
  \end{align}
 
  \begin{figure}[t!]
  \includegraphics[width=0.99\linewidth]{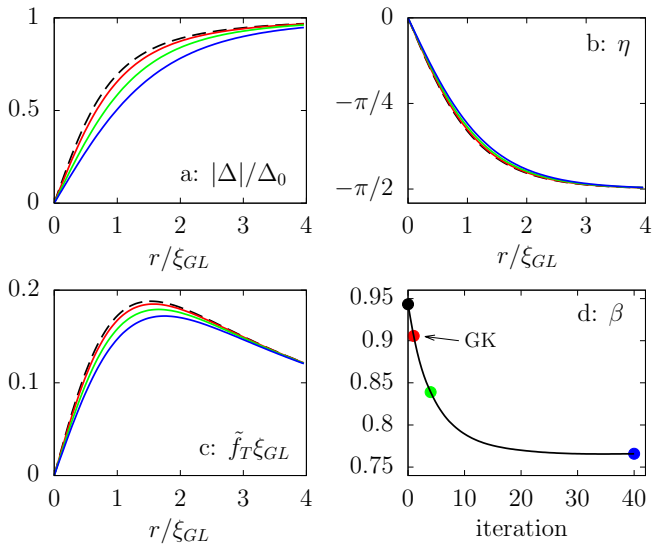}
  \caption{\label{f1} 
  (Color online) { (a) Single vortex solution of self-consistency equation
  solved by iterations at $T/T_c=0.05$. The initial guess distribution  given by the Clem ansatz (black dash)
  and first iteration (red) used in Ref. \onlinecite{GorkovKopnin1} are shown
  compared to 4th (green) and 40th (blue) iterations. In panels (b) and (c) values of the angle $\eta={\rm Im}\;\theta$ and distribution function $\tilde f_T$ are calculated based on gap profiles shown in panel (a). 
  (d) The flux-flow
  conductivity slope $\beta$ is depicted as function of iteration number.
  Values of $\beta$ which corresponds to the gap distributions shown in (a) are indicated
 by the dots with analogous colour. {Notation GK refers to the result calculated by Gorkov and Kopnin \cite{GorkovKopnin1}.}}}
  \end{figure}
  
 Previously, the value of $\beta\approx 0.9$ has been reported by {GK} \cite{GorkovKopnin1}. The calculation was based on the 
 approximate vortex solution taken from Ref. \onlinecite{watts-tobin}. 
 This vortex structure was obtained by solving iteratively the self-consistency 
 Eq. (\ref{Eq:SelfConsistency2B-2}). Each iteration step was performed as follows. First for a given vortex profile 
 the GFs at each Matsubara frequency were determined by solving the Usadel equation.
 Then these GFs were substituted to the self-consistency equation in order to calculate the updated order parameter distribution. 
 The iteration procedure used in Ref.\onlinecite{watts-tobin} started from the gap function 
 which is  known also as the Clem ansatz \cite{clem}. However, instead of taking the sufficient number of iterations to reach 
 self-consistency only a single iteration step was performed in Ref.\onlinecite{watts-tobin}. 
 In this way the approximate vortex profile was obtained which was used later to 
 calculate the flux-flow conductivity at low fields \cite{GorkovKopnin1}. 
   
 To get the correct order parameter distribution we have performed sufficient number of 
 iterations so that to ensure that the order parameter changes with each update become negligible. 
 With the help of the fully self-consistent vortex structure obtained in this way we found that $\beta=0.77$. 
 Previously reported value $0.9$ is  overestimated by $17\%$. The disparity between initial gap distribution, the one obtained after 
   the first iteration and the exact gap function together with corresponding values of $\beta$
   are shown in Fig. \ref{f1}. 
   
 \subsubsection{High-temperature limit}\label{sec:high}
 \begin{figure}[t!]
 \includegraphics[width=0.99\linewidth]{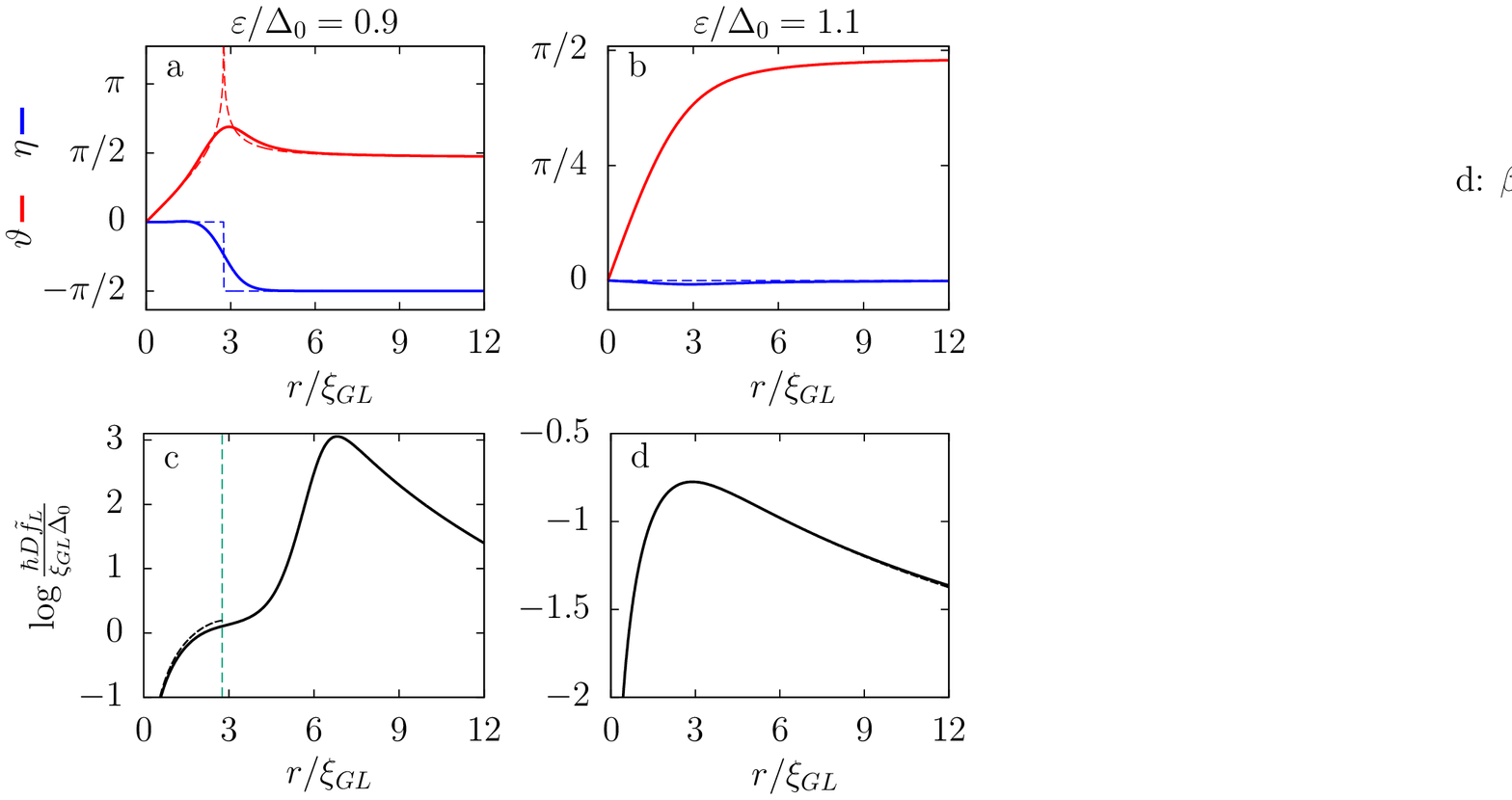}
 \caption{\label{f2} 
 (Color online)
(a,b) $\vartheta = {\rm Re}\; \theta$ (red) and $\eta = {\rm Im}\;\theta$ (blue) as functions of distance
 for the energies below (left) and above (right) the bulk gap $\Delta_0$ and temperature $T/T_c=0.99$. 
 Dashed curves represent solutions obtained within local density approximation used by LO. 
 (c,d) The logarithm of $\tilde f_L$ (solid black) for the same energies. Dashed black curve in panel (c) corresponds to the distribution function used by LO, see Eq. (\ref{tildefLsol}). Vertical line in panel (c) points to the distance where energy equals local density value, $\varepsilon=|\Delta|$. Note that for $\varepsilon>\Delta_0$ the difference between exact numerics and local density 
 approximation is rather small.
}
 \end{figure}

 For elevated temperatures but still in the diffusive-controlled limit
 $(T_c-T)\tau_{ph}\gg 1$, the non-equilibrium states are dominated mostly by the 
 change in the number of quasiparticles determined by the $f_L$ mode while the charge imbalance $f_T$ 
 yields the subdominant contribution. In this regime the  
 FFC was calculated within the local density approximation for the spectral functions\cite{kopnin-book,LarkinOvchinnikovTc2}. 
 This approximation results in the expression (\ref{EqBetaDiverge}) with $\beta_0=4.04$, see Appendix \ref{App:LO} for the calculation details.

The local density approximation is well justified in the limit $T\to T_c$.
However, to stay in the diffusion-controlled regime the temperature cannot be 
 taken infinitesimally close to the critical one.
 Thus it is interesting to improve the accuracy of $\beta$ calculation 
for small but finite values of $T_c-T$. For this purpose we find the order parameter solving  
the self-consistency equation (\ref{Eq:SelfConsistency2B-2}) numerically. 
After that we considered Eq.(\ref{Eq:theta}) for the spectral functions, where small parameter $1/\tau$ regularizes the gap edge singularities. We fixed the value $\lambda_{ph}\sim 10^{-6}$ so that relaxation time $\tau$ appears to be sufficiently large and diffusion-controlled 
FFC remains unaffected up to the temperatures $1-T/T_c\sim \lambda_{ph}$. By starting with initial distributions for $\vartheta$ and $\eta$, we calculated relaxation time $\tau$ according to Eq. (\ref{tau_theta}) and then solved numerically Eq.(\ref{Eq:theta}) to get new functions  $\vartheta$ and $\eta$. By repeating this procedure iteratively, we found spectral functions with sufficient accuracy. By using these solutions, we calculated relaxation rate $\nu_{out}$ for non-equilibrium longitudinal imbalance, Eq. (\ref{collint_out1}), and solved the kinetic equations (\ref{Eq:tildefT})-(\ref{Eq:tildefL}) by omitting scattering-in term $j_{in}$. 
Note that the relaxation term in the kinetic equation (\ref{Eq:tildefL}) allows to apply the zero boundary conditions in the bulk.

Fig. \ref{f2} demonstrates the exactly calculated $\vartheta$, $\eta$ 
 and the distribution function $\tilde f_L$ compared to those obtained within  the 
 local-density approximation. 
With these functions  we calculated integrals $\alpha$ and $\gamma$, see expression (\ref{Eq:Visc}). 
As a result, we obtained the 
divergent behaviour (\ref{EqBetaDiverge}) with the dimensionless parameter $\beta_0\approx 3.7$.
Therefore, local-density approximation overestimates $\beta_0$ by $9\%$.


\subsubsection{Intermediate temperatures}
For the temperatures within the broad range between limiting cases considered above,
the contributions of both the $f_L$ and $f_T$ modes are generically of the same order of magnitude. 
 Therefore in order to calculate the FFC it is necessary to solve system of coupled kinetic equations 
  (\ref{Eq:tildefT})-(\ref{Eq:tildefL}). This can be done only numerically, and exact temperature dependence $\beta=\beta(T)$ in the diffusion-controlled regime
 has never been calculated before. Previously only the interpolation curve between GK and LO results has been suggested\cite{kopnin-book}. 
 Below we compare this interpolation curve with the result of an exact numerical calculation which is 
 done in the same way as discussed above in Sec. \ref{sec:high} by repeating all steps at different temperatures.

  \begin{figure}[t!]
  \includegraphics[width=0.5\linewidth]{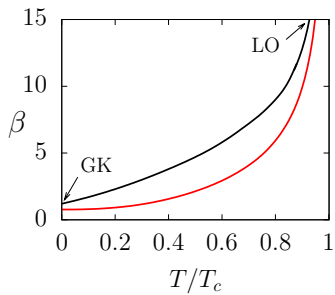}
  \caption{\label{f3} 
  (Color online) {Diffusion controlled flux-flow conductivity parameter $\beta$ (\ref{Eq:FluxFlowCondVisc}) as a function of a temperature. Black is interpolation between Gorkov-Kopnin zero-temperature value \cite{GorkovKopnin1} (GK) and Larkin-Ovchinnikov asymptote \cite{LarkinOvchinnikovTc2} (LO). 
  Red is a result of numerical calculation. 
  }}
  \end{figure}
 Shown by the red curve in Fig. \ref{f3} is the obtained temperature dependence $\beta = \beta (T)$
 which is qualitatively similar to the interpolation curve (black line).
 Both dependencies feature the gradual increase from Bardeen-Stephen limit,  $\beta \sim 1$, at small temperatures 
  to the large values of $\beta$ at high temperatures due to decrease in diffusion relaxation rate. 
  However, calculated dependence $\beta(T)$ is significantly lower compared to the interpolation 
  curve suggested previously in Ref. \onlinecite{kopnin-book}. 

 \subsection{Suppression of FFC by inelastic relaxation} \label{Sec:InelasticRelaxation}
Inelastic electron-phonon scattering provides an additional relaxation mechanism 
 which affects FFC. This relaxation channel plays an important role at temperatures 
 close to the critical one when the spatial gradients of the distribution functions 
 become small due to increase in the correlation length and superconducting energy gap is suppressed.

   \begin{figure}[t!]
   \includegraphics[width=0.99\linewidth]{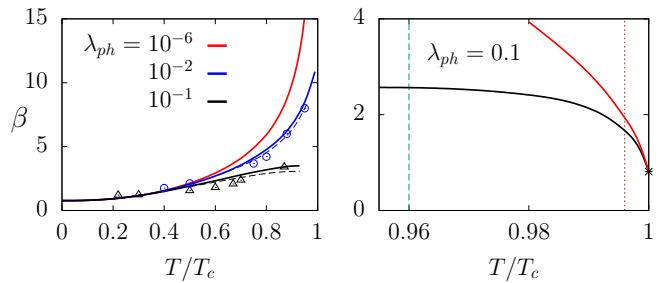}
   \caption{\label{f4} 
   (Color online) Left: FFC as electron-phonon relaxation rate determined by $\lambda_{ph}$ increases (solid curves).
   Experimental data is shown for Zr$_3$Rh \cite{poon} (blue circles) and Nb$_{0.5}$Ta$_{0.5}$ \cite{ExperimentKim,ExperimentVinen} (black triangles). Dashed curves represent FFC calculated by means of distribution functions corrected by scattering-in contribution to collision integral, see text.
   Right: FFC calculated in the limit $T\to T_c$ by neglecting non-equilibrium transverse imbalance mode (black). 
   Red curve is the Tinkham contribution to FFC in the GTDGL theory and star corresponds to its limiting value 
   in the gappless regime. Dashed vertical line is defined by condition $\hbar D/\xi_{GL}^2=T_c\lambda_{ph}$ and dotted one by 
   $\hbar D/\xi_{GL}^2=0.1T_c\lambda_{ph}$.
   }
   \end{figure}

 The crossover between diffusion-controlled and inelastic relaxation-controlled branches of the $\beta(T)$
 dependence occurs at the temperatures $\tau_{ph}( T_c -T) \sim 1 $ where none of these approximations can be applied. 
 The behaviour of $\beta (T)$ in this region of parameters has not been studied before. 
 To analyse an interplay between two relaxation regimes we calculate numerically FFC for different inelastic scattering rates 
 determined by the value of parameter $\lambda_{ph}$. We apply the same numeric procedure as was discussed in Sec. \ref{sec:high}.
 
Fig. \ref{f4} shows the result of the calculation. Inelastic electron-phonon scattering suppresses the maximal value of FFC and smears the crossover from solely diffusion-driven to inelastic relaxation controlled regimes. Such a behaviour is caused by 
suppressed generation of non-equilibrium quasiparticles due to the presence of electron-phonon relaxation channel so 
that non-equilibrium longitudinal imbalance becomes weaker.
This follows from kinetic equation (\ref{Eq:tildefL}) where electron-phonon relaxation tends to suppress the source term determined by the density gradient.   

To demonstrate the consistency of our numerics we first estimated the effect of scattering-in contribution to collision integral in kinetic equation (\ref{Eq:tildefL}). To do this we solved kinetic equations without scattering-in term for the energies in the interval $[-20\Delta_0\ldots20\Delta_0]$ and then calculated collision integral (\ref{collint_in1}). By using its value we solved kinetic equations again and obtained corrected distribution functions together with more accurate values of FFC shown in the left panel of Fig. \ref{f4} by dashed curves. The effect of scattering-in term is rather small.

Secondly, we calculated FFC as temperature approaches the critical one. In this limit the GTDGL theory\cite{watts-tobin1981} becomes of relevance. 
By neglecting electric field and charge imbalance mode we found that the remaining contribution to FFC gradually approaches the Tinkham term of the GTDGL model, see Fig. \ref{f4}. At that, the crossover towards electron-phonon relaxation controlled regime and the GTDGL theory takes place very close to the critical temperature where electron-phonon relaxation rate is at least ten times larger than the one for the diffusion. The opposite condition $\hbar D/\xi_{GL}=10T_c\lambda_{ph}$ is satisfied at the temperature $T/T_c=0.6$ for $\lambda_{ph}=0.1$ and below this limit FFC is well approximated by diffusion mechanism only, see Fig. \ref{f4}. This suggests that 
the temperature interval where FFC is characterized by coexistence of diffusion driven and inelastic scattering controlled mechanisms of relaxation can be quite wide and  none of estimations (\ref{EqBetaDiverge},\ref{Eq:BetaTDGL}) can give adequate description in this region.

The modification of FFC caused by electron-phonon scattering allows to obtain good numerical agreement with experimental data by using the inelastic relaxation rate $\lambda_{ph}=(\tau_{ph} T_c)^{-1}$ as the fitting parameter. In real superconducting systems, FFC is strongly affected by electron-phonon relaxation so that applicability of Eqs. (\ref{EqBetaDiverge}) and (\ref{Eq:BetaTDGL}) appears to be very limited. In this case, the overall temperature behaviour of FFC can be found only numerically due to multi-component mechanism of the non-equilibrium quasiparticle relaxation during the motion of the vortexes.

In Fig. \ref{f4} we compare numerically calculated curves with experimental data for Na-Ta system\cite{ExperimentKim,ExperimentVinen} and amorphous superconductor Zr$_3$Rh \cite{poon}. For the former case, good fit is achieved for the value $\lambda_{ph}=0.1$ which corresponds to the electron-phonon relaxation time about $10^{-11}$ s. This value coincide by the order of magnitude with ones reported previously for niobium \cite{gousev,ptitsina}. For the latter system, electron-phonon relaxation time is found to be about $10^{-10}$ s.
 
\section{Summary} \label{Sec:Conclusion}

To summarise, we have calculated the FFC in diffusive superconductors for small magnetic fields and arbitrary temperatures taking into account the electron-phonon relaxation and using the self-consistent vortex solutions. At first, we have obtained the exact value of the dimensionless parameter $\beta = 0.77$ which determines the FFC in the low temperature limit $T\to 0$. Second, we calculated the overall temperature dependence of $\beta$ in the domain of the diffusion-controlled flux flow, that is at $\tau_{ph}(T_c-T)\gg 1$. Significant deviations from the previously reported interpolation curve are obtained. Finally, we studied the crossover between the diffusion-controlled and generalized TDGL regimes which occurs at $\tau_{ph}(T_c-T)\sim 1$. The maximal value of $\beta$ obtained in this region is much smaller than 
   expected from the estimations based on the Eqs. (\ref{EqBetaDiverge}) and (\ref{Eq:BetaTDGL}) at the border of their applicability. 
   Consequently, we obtained significant suppression of FFC near $T_c$ by changing the electron-phonon relaxation rate and achieved better agreement with the experimental data.   
                  
 \section{Acknowlegements} 
 The work was supported by the Estonian Ministry of Education and Research (grant PUTJD141) and 
 the Academy of Finland. 
           
 \appendix

 \section{Derivation of kinetic equations} \label{App:KinDerivation}
 The quasiclassical GF matrix defined in Eq. (\ref{eq:GF0})
 obey the Usadel equation 
 \begin{equation}\label{Eq:KeldyshUsadelApp}
 \{\hat\tau_3\partial_t, \check g \}_t = D\hat\partial_{\bf r}  ( \check g
\circ \hat\partial_{\bf r} \check g) 
 + [\hat H , \check g ]_t +\check I,
  \end{equation} 
 where 
 $\hat H ({\bm r},t) = i \hat \Delta-ie\phi\hat\tau_0$, 
 $\hat\Delta (t)=  i |\Delta| \hat\tau_2e^{-i\varphi \hat\tau_3}$ 
 is the gap operator and $\check I=i(\check g\circ\check\Sigma-\check\Sigma\circ\check g)$ is collision integral due to relaxation processes described by self-energy $\check\Sigma$. 
 Equation (\ref{Eq:KeldyshUsadelApp}) is complemented by the normalization condition and the parametrization
 of the Keldysh component is introduced in (\ref{Eq:Parametrization}). 
 Throughout the derivation we assume $k_B=\hbar=c=1$.

The diagonal elements of matrix equation (\ref{Eq:KeldyshUsadelApp}) give equations for $\hat g^{R/A}$ which have 
same form as (\ref{Eq:KeldyshUsadelApp}) with 
$\hat I^{R/A}=i(\hat g^{R/A}\circ\hat\Sigma^{R/A}-\hat\Sigma^{R/A}\circ\hat g^{R/A})$ 
substituted. The non-diagonal element reads as
 \begin{align}\label{Eq:KeldyshUsadelKeldyshComponent}
 &\{\hat\tau_3\partial_t,  \hat g^K \}_t = D\hat\partial_{\bf r} ( \check g \circ \hat\partial_{\bf r} \check g)^K+[\hat H, \hat g^K]_t+\hat I^K,
 \end{align}   
 where
 \begin{align}
 &\hat I^K=i(\hat g^R\circ\hat \Sigma^K+\hat g^K\circ\hat \Sigma^A-\hat \Sigma^R\circ\hat g^K-\hat \Sigma^K\circ\hat g^A),
 \nonumber\\
 &\hat\partial_{\bf r} ( \check g \circ \hat\partial_{\bf r} \check g)^K =
  \hat\partial_{\bf r} ( \hat g^R  \circ \hat\partial_{\bf r} \hat g^K + \hat g^K \circ \hat\partial_{\bf r} \hat g^A  )= 
  \\
  &\hat\partial_{\bf r} (\hat\partial_{\bf r} \hat f - \hat g^R\circ\hat\partial_{\bf r}\hat f\circ \hat g^A )+
   \hat g^R\circ\hat\partial_{\bf r} \hat g^R\circ \hat\partial_{\bf r} \hat f - 
   \nonumber\\
  &\hat\partial_{\bf r} \hat f\circ \hat g^A\circ\hat\partial_{\bf r} \hat g^A +  
  \hat\partial_{\bf r} (\hat g^R\circ\hat\partial_{\bf r} \hat g^R )\circ \hat f - \hat f \circ \hat\partial_{\bf r}
  (\hat g^A\circ\hat\partial_{\bf r} \hat g^A ). 
  \nonumber
 \end{align}
 To obtain last relation we substituted parametrization (\ref{Eq:Parametrization}) and used the associative property of differential
superoperator $ \hat\partial_{\bf r}(g_1\circ g_2)= \hat\partial_{\bf r}g_1\circ g_2 + g_1\circ\hat\partial_{\bf r} g_2$.  To get rid of the last two terms we subtract the spectral components of the
Eq.(\ref{Eq:KeldyshUsadelApp}) to obtain finally
 the equation
  \begin{eqnarray} \label{Eq:KineticEqGen}
  & \hat g^R \circ\{\hat\tau_3 \partial_{t},\hat f\}_t -\{\hat\tau_3\partial_{t},\hat f\}_t\circ\hat g^A=
  \nonumber \\
  & D\hat\partial_{\bm r} (\hat\partial_{\bm r} \hat f - \hat g^R\circ \hat\partial_{\bm r}\hat f\circ \hat g^A )+
  \nonumber\\ 
  & D ( \hat g^R\circ\hat\partial_{\bf r} \hat g^R\circ \hat\partial_{\bm r} \hat f- 
   \hat\partial_{\bm r} \hat f\circ \hat g^A\circ\hat\partial_{\bf r} \hat g^A )+
   \nonumber\\
   &\hat g^R \circ [\hat H,\hat f]_t- [\hat H,\hat f]_t\circ\hat g^A +\hat J,
  \end{eqnarray}
where $\hat J=\hat Z\circ \hat g^A-\hat g^R\circ\hat Z$ and $\hat Z=i(\hat \Sigma^R\circ\hat f-\hat f \circ\hat \Sigma^{A}-\hat\Sigma^K)$. In our consideration, collision integral describes electron-phonon scattering channel only. This term is responsible for establishing the equilibrium in the system.

  To proceed we introduce the mixed representation in the time-energy domain 
  as follows
  $\hat g (t_1,t_2) = \int_{-\infty}^{\infty} \hat g(\varepsilon, t) 
  e^{-i\varepsilon (t_1-t_2) } \frac{d\varepsilon}{2\pi} $, 
  where $t=(t_1+t_2)/2$. To keep the gauge invariance we introduce the modified GF 
 $\hat g_{new}  = \hat W(t_1, t) \hat g (t_1,t_2) \hat W(t,t_2)$ where the link operator is defined in the text. 
 This transformation removes the scalar potential term from the kinetic equations and adds the 
 chemical potential shift $e\phi$. We absorb this shift by substituting 
$f_T \to f_T +e \phi \partial_\varepsilon f_0$, where $f_0(\varepsilon)=\tanh[\varepsilon/(2T)]$ is equilibrium distribution, 
so that $f_T$ hereafter denotes the deviation from the local equilibrium. 
 Then keeping the first order terms in frequency, we get the gradient approximation
  \begin{align}
 & [\hat H , \hat g ]_t = [\hat H , \hat g ] - i\{ \partial_t \hat H, \partial_\varepsilon\hat g\}/2, 
 \nonumber\\
 & [\bm A \hat\tau_3, \hat g ]_t = \bm A[ \hat\tau_3, \hat g ] - i \partial_t{\bm A} \{ \hat\tau_3, \partial_\varepsilon\hat g \}/2, 
  \nonumber\\\label{A9}
 & \hat\partial_{\bf r} {\hat f} (\varepsilon,t)  = \nabla ( f_L\hat\tau_0 + f_T \hat\tau_3) + e{\bm E} \hat\tau_3 \partial_\varepsilon f_0,
  \end{align}
 where ${\bm E}= - \nabla \phi -\partial_t{\bm A}$ is electric field.
 
    Here we assume the first order in deviation from equilibrium so that equilibrium distribution 
 $f_0$ is substituted in last term in (\ref{A9}). With the same accuracy we obtain
 \begin{align} \label{ddf}
 &\hat\partial_{\bf r} (\hat\partial_{\bf r} \hat f - \hat g^R\circ
 \hat\partial_{\bf r}\hat f\circ \hat g^A )  = 
 \\ \nonumber
 & \hat\nabla( \hat \nabla \hat f_1 - \hat g^R \hat \nabla\hat f_1 \hat g^A) + 
 e \partial_\varepsilon f_0 \nabla \cdot ({\bm E} (\hat\tau_3-\hat g^R \hat\tau_3\hat g^A)) 
 \end{align}         
   where $\hat f_1  = (f_L - f_0)\hat\tau_0 +  \hat\tau_3 f_T$ is the deviation from the local equilibrium and  $\hat\nabla = \nabla\hat\tau_0 - ie {\bm A} [\hat\tau_3, ]$ is gauge-covariant gradient. In (\ref{ddf}) we keep only terms which contribute to the kinetic equations.


In the mixed representation the kinetic Eq.(\ref{Eq:KineticEqGen}) has the
following gauge-invariant form
  \begin{eqnarray} \label{Eq:KineticEq}
 &  D\nabla(\nabla \hat f_1 - \hat g^R \nabla \hat f_1 \hat g^A) +
 \\ &
  D( \hat g^R\hat\nabla \hat g^R\nabla \hat f_1 - 
  \nabla \hat f_1 \hat g^A\hat\nabla \hat g^A) +  \nonumber\\
  &\hat g^R [\hat H, \hat f_1] - [\hat H, \hat f_1]\hat g^A - i \partial_\varepsilon f_0 
  ( \hat g^R \hat\partial_t \hat H  - \hat \partial_t \hat H \hat g^A)  + 
  \nonumber\\ \nonumber
  & eD \partial_\varepsilon f_0  \nabla \cdot 
  \left( {\bm E} (\hat\tau_3- \hat g^R \hat\tau_3\hat g^A)\right) + 
 \\ \nonumber
 & eD \partial_\varepsilon f_0 {\bm E}\cdot ( \hat g^R\hat\nabla
 \hat g^R\hat\tau_3 - \hat\tau_3 \hat g^A \hat\nabla \hat g^A)  +  \hat J
 =0.
 \end{eqnarray}
  Here we took into account only first-order terms in the deviation from equilibrium and introduced gauge-covariant time derivative $\hat \partial_t  = \hat\tau_0\partial_t + 2ie\phi \hat\tau_3$. 
 
    To obtain the equations (\ref{Eq:KineticEqFT3}) and (\ref{Eq:KineticEqFL}) in the main text  
  we trace Eq. (\ref{Eq:KineticEq}) with Nambu matrices $\hat\tau_0$ and $\hat\tau_3$ respectively. Here we took into account that ${\rm Tr} ( \hat g^R \hat\tau_3\hat g^A) =0$
    because of the relation $\hat g^A= -\hat\tau_3 \hat g^{R+} \hat\tau_3$
    and the general form of the equilibrium spectral function $\hat
  g^R=g_3\hat\tau_3 + g_2\hat\tau_2 e^{-i\varphi\hat\tau_3}$. Then we neglect the driving terms with electric field  and electron-phonon relaxation of the charge  
  imbalance to get the Eq.(\ref{Eq:KineticEqFT3}). We keep the electron-phonon collision integral in the 
  Eq.(\ref{Eq:KineticEqFL}) which plays an important role in vortex dynamics being the only energy relaxation channel.
 

\section{Collision integrals}\label{App:coll}

We consider small non-stationary corrections to the GF in the form $\hat g^{R/A}=\hat g^{R/A}+\hat g^{R/A}_{nst}$ and $\hat g^K=\hat g^{RA}f_0+\hat g^{nst}$, where $\hat g^{R/A}\gg\hat g^{R/A}_{nst}$ and $\hat g^{nst}$ defined in Eq. (\ref{Eq:GnstExp}). Here we use notation $X^{RA}=X^R-X^A$ for $X=\hat g$. Then the stationary parts of inelastic electron-phonon self-energy (\ref{Eq:ElectronPhononSE}) 
 read as
 \begin{align}\label{tildesigma}
  & \tilde \Sigma^{R/A} = 2i\omega|\omega|\bigg\{ \hat g^{R/A}(\varepsilon + \omega)\left[\frac{1}{f_0(\omega)}-f_0(\varepsilon + \omega)\right]+\nonumber\\
  &\qquad\qquad \frac{f_0(\varepsilon + \omega)}{2}\left[\hat g^{R}(\varepsilon + \omega)+\hat g^A(\varepsilon + \omega)\right] \bigg\},
  \\
 & \tilde\Sigma^K = 2i\omega|\omega| \hat g^{RA}(\varepsilon+\omega)
 \left[ f_0(\varepsilon+\omega)/f_0(\omega)- 1 \right].  \nonumber
\end{align}
We are mostly interested in the self-energies at $\varepsilon\sim\Delta_0$, while the dominant contribution to the integral (\ref{Eq:ElectronPhononSE}) is coming from the region $\omega\gg\Delta_0$. 
Since for higher energies $\hat g^R+\hat g^A\ll \hat g^{R/A}$ and $\hat g^{R/A}\approx\hat \tau_3 g^{R/A}$, the second contribution to $\tilde \Sigma^{R/A}$ in Eq. (\ref{tildesigma}) can be neglected and
the self energy can be presented in the relaxation-time approximation, $i\hat\Sigma^{R/A}=\pm\hat\tau_3/(2\tau)$, where  
$\tau$ is energy-dependent inelastic electron-phonon collision time defined by
\begin{align}\label{tau}
\frac{1}{\tau} =\frac{\lambda_{ph}\cosh\frac{\varepsilon}{2T}}{14\zeta(3)T_c^2} 
\int_{-\infty}^{\infty} \frac{ \omega|\omega| d\omega g^R(\varepsilon+\omega) }{\sinh\frac{\omega}{2T} \cosh\frac{\varepsilon + \omega}{2T}}.
\end{align}
This expression coincides with the formula used by Watts-Tobin {\it et al.} \cite{watts-tobin1981} In Eq. (\ref{tau}), relaxation time $\tau$ can contain imaginary part due to complex $g^R$. Usually this contribution is absorbed by the renormalizing of the chemical potential. Note that near critical temperature, where $g^R\approx1$ and $\varepsilon\ll\omega\sim T_c$, inelastic electron-phonon collision time approaches value
 $\tau_{ph}=1/(\lambda_{ph}T_c)$. 

Next, we express non-stationary contributions to self-energies via $\hat g^{R/A}_{nst}$ and $\hat g^{nst}$ and derive the mixed representation for $\hat Z=i(\hat \Sigma^R\circ\hat f-\hat f \circ\hat \Sigma^{A}-\hat\Sigma^K)$. The latter quantity does not contain stationary terms. For collision integral 
$\hat J=\hat Z\circ \hat g^A-\hat g^R\circ\hat Z$ in the mixed representation we obtain with the help of GF in the Nambu space ${\rm Tr}\hat J=-(f_L-f_0)\nu_{out}+J_{in}$, where
\begin{align}\label{collint_out}
&\nu_{out}=\frac{\lambda_{ph}}{28\zeta(3)T_c^2}\int_{-\infty}^\infty d\omega\omega|\omega|\Big\{2g^{RA}(\varepsilon)g^{RA}(\varepsilon+\omega)-\nonumber\\
&f^{RA}(\varepsilon)f^{R+A+}(\varepsilon+\omega)-f^{R+A+}(\varepsilon)f^{RA}(\varepsilon+\omega)\Big\}\times\nonumber\\
&\left[1/f_0(\omega)-f_0(\varepsilon + \omega)\right],\\
&J_{in}=\frac{\lambda_{ph}}{28\zeta(3)T_c^2}\int_{-\infty}^\infty d\omega\omega|\omega|\Big\{2g^{RA}(\varepsilon)g^{RA}(\varepsilon+\omega)-\nonumber\\
&f^{RA}(\varepsilon)f^{R+A+}(\varepsilon+\omega)-f^{R+A+}(\varepsilon)f^{RA}(\varepsilon+\omega)\Big\}\times\nonumber\\
&\left[f_L(\varepsilon + \omega)-f_0(\varepsilon + \omega)\right][f_0(\varepsilon)+1/f_0(\omega)]. \label{collint_in}
\end{align}
Here we used notation $f^{R+A+}=f^{R+}-f^{A+}$. In the expressions (\ref{collint_out}) and (\ref{collint_in}) the dominant contribution to the integrals is coming from high energies. 
Since $f_L-f_0$ is significant only at low energies, 
scattering-in term $J_{in}$ appears to be a small correction to the collision integral. 
Note that $\nu_{out}\to2g^{RA}/\tau$, if temperature approaches critical one.

 \section{$\theta$-parametrization} \label{App:theta}           
 
The Usadel equation for equilibrium spectral functions has the form
 \begin{align}\label{usadelspectral}
 D\hat\partial_{\bf r}(\hat g^{R/A}\hat \partial_{\bf r}\hat g^{R/A})+[i\varepsilon \hat\tau_3+i\hat \Delta-i\hat\Sigma^{R/A},\hat g^{R/A}]=0.
 \end{align}
By deriving this equation, we took into account that in the mixed representation $\{\hat\tau_3\partial_t,\hat g\}_t = -i\varepsilon[\hat\tau_3,\hat g]$ and scalar potential is neglected in the equilibrium.
 
By using parametrization (\ref{Eq:SpectralGF-R}, \ref{Eq:SpectralGF-A}) one finds in cylindrical coordinates
\begin{align}
& [\hat \Delta+\varepsilon\hat\tau_3+i\hat\tau_3/(2\tau),\hat g^R] = 
 \nonumber \\
& 2\{[\varepsilon+i/(2\tau)]\sinh\theta - |\Delta|\cosh\theta\}\hat\tau_1 e^{-i\hat\tau_3\varphi}, 
\nonumber \\
& \nabla(\hat g^R\nabla\hat g^R)=\left[\nabla_r^2\theta-\sinh(2\theta)/(2r^2)\right]\hat\tau_1e^{-i\hat\tau_3\varphi},
\end{align}
where $\nabla_r^2=\partial_r^2+r^{-1}\partial_r$. By taking into account that self-energy $\hat \Sigma^{R/A}$ in Eq. (\ref{usadelspectral}) corresponds to the stationary contribution, $2i\hat \Sigma^{R/A}=\pm \hat\tau_3/\tau$ (see Appendix \ref{App:coll}),
 %
one obtains Eq. (\ref{Eq:theta}) for $\theta(r)$. 

It is convenient to split $\theta$ into the real and imaginary parts, $\vartheta  ={\rm Re } \theta $ and $\eta = {\rm Im}\theta$, which satisfy the following equations
 \begin{align}\label{Eq:SpectralGFcyl}
 & \nabla^2_r \vartheta - \frac{ \sinh(2\vartheta)\cos(2\eta)}{2r^2} =\\
 & \qquad 
 \frac{2}{D}\left(\varepsilon\cosh\vartheta\sin\eta + 
 \frac{1}{2\tau} \sinh\vartheta\cos\eta - |\Delta|\sinh\vartheta\sin\eta \right), 
 \nonumber
 \\
 &\nabla^2_r \eta - \frac{ \cosh(2\vartheta)\sin(2\eta)}{ 2r^2 }=
 \nonumber
 \\
 &\qquad \frac{2}{D} \left( |\Delta| \cosh\vartheta \cos\eta - 
 \varepsilon \sinh\vartheta \cos\eta + 
 \frac{1}{2\tau}\cosh\vartheta \sin\eta \right),
 \nonumber
 \end{align}
supplemented by the boundary conditions (\ref{thetabc}). 

With the help of parametrization (\ref{Eq:SpectralGF-R},\ref{Eq:SpectralGF-A}), kinetic equation can be simplified due to the following
identities 
   \begin{align}
 & {\cal D}_T =2D[ 1+\cosh(2\vartheta) ],\nonumber\\
 & 2i {\rm Tr} [ (\hat g^R + \hat g^A) \hat \Delta] = 8 |\Delta|\cosh\vartheta\sin\eta, 
 \nonumber\\ 
 & {\rm Tr} [ \hat\tau_3 \partial_t\hat \Delta( \hat g^R+ \hat g^A) ] =
 -4({\bm v}_L\cdot \nabla\varphi) |\Delta| \cosh\vartheta\sin\eta,
 \nonumber\\
 & {\cal D}_L=2D[1 + \cos(2\eta )], \\
 & {\rm Tr}[\hat\tau_3(\hat g^R-\hat g^A)\hat \Delta]=0,
 \nonumber\\
 & {\rm Tr}[\partial_t\hat \Delta(\hat g^R-\hat g^A)] = 
 4({\bm v}_L \cdot\nabla|\Delta|)\sinh\vartheta\cos\eta,
 \nonumber\\
 &{\bm j}_e = -2D\sinh (2\vartheta) \sin( 2\eta) \nabla\varphi,
 \nonumber
 \end{align}
  where we took into account that for the vortex moving with constant velocity
$\partial_t\Delta=- {\bm v}_L\cdot\nabla\Delta$. By construction, spectral current is conserved, $\bm\nabla\cdot{\bm j}_e=0$.
 Taking into account that ${\bm v}_L\cdot\nabla\varphi =-v_L\sin\varphi /r$
 and ${\bm v}_L\cdot\nabla|\Delta| = v_L \cos\varphi \partial_r |\Delta|$ 
 we arrive to Eq. (\ref{Eq:tildefT})- (\ref{Eq:tildefL}), where collision integral ${\rm Tr}\hat J=-(f_L-f_0)\nu_{out}+J_{in}$ (see Appendix \ref{App:coll}) is substituted. At that, we used $\theta$-parametrization to obtain
 \begin{align} 
 &2g^{RA}(\varepsilon)g^{RA}(\varepsilon^\prime)-f^{RA}(\varepsilon)f^{R+A+}(\varepsilon^\prime)-f^{R+A+}(\varepsilon)f^{RA}(\varepsilon^\prime)\nonumber\\
 &=8\cos[\eta(\varepsilon)]\cos[\eta(\varepsilon^\prime)]\cosh[\vartheta(\varepsilon)-\vartheta(\varepsilon^\prime)],
 \end{align}
 and renormalized scattering-in part, namely $J_{in}=v_{L}j_{in}\cos\varphi \partial_\varepsilon f_0$.
   
 To calculate the force ${\bm F}_{env}$ (\ref{Eq:FenvGen}) we use the expansion (\ref{Eq:GnstExp}) and  the 
 spectral functions in the form (\ref{Eq:SpectralGF-R}, \ref{Eq:SpectralGF-A}).
 Using the ansatz (\ref{Eq:DFansatz}), we get an expression for the force in the form
 \begin{align} \label{} 
  {\bm F}_{env} =&\ \frac{\nu v_L}{2}\int d^2 {\bm r} d\varepsilon \partial_\varepsilon f_0 
  \Big\{ \\
& \sin\varphi\cosh\vartheta\sin\eta|\Delta|\left(2\tilde f_T-1/r\right)\nabla\varphi +\nonumber\\
& \cos\varphi \Big[\partial_r(\cosh\vartheta\sin\eta)-2\tilde f_L\sinh\vartheta\cos\eta
\Big]\nabla|\Delta|\Big\}.
\nonumber   
 \end{align}   
After integration, this can be written as ${\bm F}_{env} = -\varrho {\bm v}_L$, where the viscosity coefficient is given by $\varrho= \pi\hbar\nu (\alpha+ \gamma)$ and
 \begin{align} \label{Eq:Visc}
 & \alpha =\\
 & \int\limits_0^\infty rdr \partial_r|\Delta|\int\limits_0^\infty d\varepsilon \partial_\varepsilon f_0 [2\tilde f_L\sinh\vartheta\cos\eta-\partial_r(\cosh\vartheta\sin\eta)],   \nonumber\\
 & \gamma = \int_0^\infty dr |\Delta|\int_0^\infty d\varepsilon \partial_\varepsilon f_0 \cosh\vartheta\sin\eta(2\tilde f_T-1/r).
 \nonumber
 \end{align}
Here we have took into account that $\tilde f_T$ and $\eta$ are even, while $\tilde f_L$ and $\vartheta$ are odd functions of energy $\varepsilon$.

\section{Derivation of the LO result} \label{App:LO}
Following LO \cite{LarkinOvchinnikovTc2}, analytical result for diffusion-driven FFC can be obtained by noticing that near $T_c$ the diffusion terms in the Usadel equation (\ref{Eq:theta}) are much smaller than the gap field. As a result, local density approximation can be implemented, where $\vartheta$ and $\eta$ are determined by their homogeneous expressions with bulk gap substituted by local value of gap field. 
 
To calculate conductivity contributions (\ref{Eq:Visc}), it is convenient to consider energetic integration in domains $[0\ldots|\Delta(r)|]$ and $[|\Delta(r)|\ldots\infty]$ separately. The former gives negligible contribution close to $T_c$ and can be omitted. In the latter case, energetic integration variable exceeds local gap value and local approximation results in $\eta=0$ and $\sinh\vartheta=|\Delta(r)|/\sqrt{\varepsilon^2-|\Delta(r)|^2}$. In this case, $\tilde f_T=\gamma=0$ and kinetic equation (\ref{Eq:tildefL}) is satisfied by the solution \cite{LarkinOvchinnikovTc2}
 \begin{align} \label{tildefLsol}
    &\tilde f_L= \frac{1}{r\hbar D}\int_0^r dr_1 r_1\left(\sqrt{\varepsilon^2-|\Delta|^2}-C(\varepsilon)\right), 
    \end{align}
Condition for vanishing heat current $\partial_r\tilde f_L=0$ in the bulk defines constant $C=\sqrt{\varepsilon^2-\Delta_0^2}\equiv C_1$ at large energies $\varepsilon>\Delta_0$.
 For sub-gap region $\varepsilon<\Delta_0$, LO used boundary condition with zero heat current at the interface $r=r_\varepsilon$ defined  by $\varepsilon=|\Delta(r_\varepsilon)|$. This determines integration constant for $\varepsilon<\Delta_0$ in the form $C=-\frac{2}{r_\varepsilon^2}\int_0^{r_\varepsilon} rdr\sqrt{\varepsilon^2-|\Delta(r)|^2}\equiv C_2$.

Dominant contribution to viscosity (\ref{Eq:Visc}) is stemming from integration over $\varepsilon-r$ domain enclosed by $r=0$ and $\varepsilon=|\Delta(r)|$ curves. One obtains $\alpha=\alpha_1+\alpha_2$, where
 \begin{align} \label{alpha}
     & \alpha_1=\frac{2}{\hbar D}  \int_{\Delta_0}^\infty d\varepsilon\partial_\varepsilon f_0\int_0^\infty dr r \left(\sqrt{\varepsilon^2-|\Delta|^2}-C_1(\varepsilon)\right)^2,\nonumber\\ 
    & \alpha_2=\frac{2}{\hbar D}\int_0^{\Delta_0} d\varepsilon\partial_\varepsilon f_0\left[\int_0^{r_\varepsilon} dr r (\varepsilon^2-|\Delta|^2)+\frac{r_\varepsilon^2}{2}C_2(\varepsilon)^2\right].
  \end{align}
By finding gap profile near $T_c$ numerically, we calculated these integrals and obtained $\alpha_{1,2}\approx (0.409;0.496)\xi_{GL}^2\Delta_0^3/(\hbar DT_c)$. As a result, $\beta=\beta_0\sqrt{T_c/(T_c-T)}$, where $\beta_0=4.01$. 

\bibliography{FluxFlow_exc}

 \end{document}